\documentclass[nonacm,sigconf,10pt]{acmart}

\usepackage{graphicx}
\usepackage{color}
\usepackage{hyperref}
\usepackage{booktabs}
\usepackage{multirow}
\usepackage[lined]{algorithm2e}
\usepackage{balance}
\usepackage[utf8]{inputenc}
\usepackage{amsmath}
\usepackage{wrapfig}
\usepackage{caption}
\usepackage{balance}

\renewcommand{\epsilon}{\varepsilon}

\newcommand{\BT}[1]{{\Theta}\left(#1\right)}

\newcommand{\simpl}[1]{\operatorname{simpl}(#1)}

\providecommand{\brc}[1]{\left\{ {#1} \right\}}

\renewcommand{\Re}{{\rm I\!\hspace{-0.025em} R}}
\newcommand{\Na}{{\rm I\!\hspace{-0.025em} N}}

\newcommand{\Curves}[1]{\ensuremath{\Delta^{#1}}\xspace}

\newcommand{\fresh}{FRESH\xspace}

\newcommand{\X}{\mathcal{X}} 

\begin{document}
\title{FRESH: Fréchet Similarity with Hashing}

\author{Matteo Ceccarello}
\orcid{0000-0003-2783-0218}
\affiliation{
  \institution{IT University}
  \city{Copenhagen}
  \state{Denmark}
}
\email{mcec@itu.dk}

\author{Anne Driemel}
\orcid{}
\affiliation{
  \institution{University of Bonn}
  \city{Bonn}
  \state{Germany}
}
\email{driemel@cs.uni-bonn.de}

\author{Francesco Silvestri}
\orcid{0000-0002-9077-9921}
\affiliation{
  \institution{University of Padova}
  \city{Padova}
  \state{Italy}
}
\email{silvestri@dei.unipd.it}

\maketitle

\begin{abstract}
This paper studies the $r$-range search problem for curves under the continuous Fréchet distance: given a dataset $S$ of $n$ polygonal curves and a threshold $r>0$, construct a data structure that, for any query curve $q$, efficiently returns all entries in $S$ with distance at most $r$ from $q$.
We propose \fresh, an approximate and randomized approach for $r$-range search, that leverages on a locality sensitive hashing scheme for detecting candidate near neighbors of the query curve, and on a subsequent pruning step based on a cascade of curve simplifications. 
We experimentally compare \fresh to exact and deterministic solutions, and we show that high performance can be reached by suitably relaxing precision and recall.

\keywords{Similarity search\and range reporting\and locality sensitive hashing\and Fréchet distance \and algorithm engineering}
\end{abstract}

\section{Introduction}\label{sec:intro}
The target of this paper is similarity search for time series and trajectories or, more generally, for curves: indeed, time series and trajectories can  be envisioned as polygonal curves with vertices from $\Re^d$, for a suitable dimension $d\geq 1$.\footnote{Usually, we have $d=1$ for time series and $d > 1$ for trajectories.}.
Similarity search of curves frequently arises in several applications, like
ridesharing recommendation~\cite{Shang2014},
frequent routes~\cite{Luo13},
players performance~\cite{Gudmundsson17}, and seismology~\cite{Rong18}.
In the paper, we address the \emph{$r$-range search problem}: given a dataset $S$ of $n$ curves from a domain $\X$ and a threshold $r>0$, construct a data structure that, for any query curve $q\in \X$, efficiently returns \emph{all entries} in $S$ with distance at most $r$ from $q$.
Range reporting is a primitive widely used for solving the similarity join  and   $k$-nearest neighbor problems.
 
There is no common agreement on the best distance measure for curves, for it depends on the application domain, quality of input data, and performance requirements.
There are several functions to measure the distance between two curves, such as continuous Fréchet distance, Dynamic Time Warping (DTW), Euclidean distance, and Hausdorff distance. 
We focus on the  \emph{continuous Fréchet distance}, that was introduced in computer science by Alt and Godau in the '90s~\cite{AltGodau95}.
The continuous Fréchet distance and its discrete variant, named discrete Fréchet distance \cite{Eiter94}, have been widely studied in theory (e.g. \cite{indyk2002,AfshaniD18,bm-adfd-16}) and used in different applications, like handwriting recognition~\cite{Sriraghavendra07}, protein structure alignment~\cite{Wylie13} and, in particular, trajectories of moving objects (e.g.,~\cite{Konzack17}). 
Recently, the Fréchet distance has been  addressed by the ACM SIGSPATIAL Cup 2017, drawing attention to this measure from a practical domain.

The Fréchet distance\footnote{If not differently stated, ‘‘Fréchet distance" refers to the continuous definition.} between two curves is traditionally explained with this metaphor: a man is walking on a curve and his dog on another curve; the man and dog follow their curves from start to end and can vary their speeds, but they cannot go backward; the minimum length of the leash necessary to connect man and dog during the walk is the continuous Fréchet distance. 
The Fréchet distance does not require a one-to-one mapping between points of two curves, and it is hence invariant under differences in speed: this allows, for instance, to detect the trajectories of two cars following the same street but with different speeds due to traffic conditions.

Range search is known to be computational demanding  in high dimensions under different distances, including the Fréchet distance: from a worst-case point of view, there is indeed evidence that it is not possible to obtain a truly sublinear algorithm  unless with a breakthrough for the Satisfiability problem \cite{bm-adfd-16,driemel_locality-sensitive_2017}.
Locality Sensitive Hashing (LSH), introduced in \cite{IndykM98},  is  the most common technique for  developing approximate and randomized algorithms for similarity search problems. 
LSH is a hashing scheme where near points have a higher collision probability than far points. 
Recently, \cite{driemel_locality-sensitive_2017} has introduced a family of  LSH  schemes for curves under the discrete Fréchet and  Dynamic Time Warping distances.

\subsection{Our results}
The goal of this paper is to describe and experimentally evaluate \fresh, an approximate and randomized approach for  $r$-range search  under the continuous Fréchet distance.
\fresh  builds on the theoretical ideas in \cite{driemel_locality-sensitive_2017} and extends it by providing a solid and efficient framework for trading  precision and performance. 

{\bf Algorithm design.} The core component of \fresh is a filter based on the LSH scheme for the discrete Fréchet distance in~\cite{driemel_locality-sensitive_2017}, which is boosted with  multiply-shift hashing~\cite{Dietzfelbinger97} and  tensoring~\cite{andoni_efficient_2006,christiani_fast_2017}  for better performance. 
For a given input set $S$ with $n$ curves and a query curve $q$, the filter selects as candidate near neighbors all curves colliding with $q$ under at least one of $L$  hash functions randomly selected from the LSH scheme.
This filters out a significant number of curves, without even reading them. 
All candidates are associated with a \emph{score}, representing the fraction of collisions under the $L$ hash functions.
If \fresh is seen as a classifier for detecting near and far curves for a given query $q$,  the score of a curve $p$ represents the probability that $p$ and $q$ are near.

The second component of \fresh is a candidate pruning step for reducing false positives (i.e., far curves marked as near).
The pruning consists in verifying that the fraction $0\leq \tau\leq 1$  of  candidates with smaller scores have continuous Fréchet distance from the query not larger than $r$.
As verifying the Fréchet distance is a costly operation, we propose a procedure exploiting a cascade of curve simplifications from~\cite{DriemelHW12} and verification heuristics  from~\cite{BaldusB2017,BuchinDDM17}: each step can successfully show that the distance is larger or not than $r$, or it can fail and do not provide an answer; the procedure applies the aforementioned simplifications and heuristics until one of them succeeds.

{\bf Performance/quality trade-off.} \fresh trades the quality of the results with the overall performance by suitably settings the aforementioned $L$ and $\tau$ parameters.\footnote{In addition to parameters $\tau$ and $L$, the \fresh algorithm has other second order parameters that are introduced  in Section \ref{sec:algo}, which marginally affect performance and quality. However, from an application point of view, the trade-off is mainly captured by $L$ and $\tau$, and the remaining parameters can be left to the default value in the implementation.}:
We measure the quality of the results  in terms of: 1) \emph{recall}, that is the fraction of true positives reported by the algorithm over all the positives in the ground truth; 2) \emph{precision}, that is the fraction of true positives over the predicted positives (i.e. the sum of true positives and false positives).
By increasing the number $L$ of hash functions used in \fresh, it is possible to increase the recall   of our algorithm by increasing the query time  (linear in $L$) and of the space requirements (equal to $L\cdot n+I$, where $I$ is the input size).
Once the recall has been fixed, it is possible to improve the precision  by increasing the $\tau$ parameter at the cost of a higher query time. 
The recall is not affected by this step and a perfect precision is reached by setting $\tau=1$.

{\bf Practical and theoretical guarantees.} We have carried out an extensive experimental evaluation of the \fresh algorithm over several datasets.
To evaluate \fresh, we use it as a primitive for solving a self-similarity join on each dataset $D$: specifically, for every curve in $D$, we perform an $r$-range search query  over $D$.
The experiments show that the scores computed under a query $q$ provide a good indicator of the distance from $q$, and thus filtering points according with scores is a sound approach.
From a performance point of view, we compare \fresh with the exact solutions that won the ACM SIGSPATIAL 2017 challenge~\cite{BaldusB2017,BuchinDDM17,DutschV17}. 
When the recall is approximately 70-80\% and the precision is approximately  50\%, \fresh exhibits better running times  with speedups above 5x for some inputs. Although the precision is low, the returned points are never too far from the query (up to a constant factor from $r$) by the property of the LSH scheme.
With higher precision, the heuristics adopted in the exact solutions, in particular the bounding box   approach in \cite{DutschV17}, are very effective with the 1-dimensional datasets (i.e., time series) considered in the experiments and highlight the limitations of \fresh in this setting.
\fresh is also supported by the theoretical foundations of the LSH scheme in~\cite{driemel_locality-sensitive_2017}.

The \fresh algorithm is described in Section \ref{sec:algo} and the experimental results in Section \ref{sec:exp}. The code of \fresh  is  available at
\url{https://github.com/Cecca/FRESH}.
We refer to the full version~\cite{fullversion} for a more detailed coverage of our results, including the theoretical analysis bounding the collision probability and further experiments.

\subsection{Related works}

{\bf Similarity search for curves.} Data structures for searching among curves  under the  Fréchet distance  have been studied under different angles. 
One of the earlier theoretical works is~\cite{indyk2002} that proposes a nearest neighbor data structure for Fr\'echet distance. In 2011, \cite{bcg-ffq-11} revived the topic  motivated by the availability of high-resolution trajectories of soccer players in the emerging area of sports analytics. A comprehensive study of the complexity of range searching under the Fr\'echet distance appeared in~\cite{AfshaniD18}, that also gives lower bounds on the space-query-time trade-off of range searching under the Fr\'echet distance. 
Recently, the annual data competition within the ACM SIGSPATIAL conference on geographic information science has drawn attention to the timeliness of this problem~\cite{werner2018acm}. 
The focus of the challenge was on exact solutions and hence none of the awarded submissions~\cite{BaldusB2017,BuchinDDM17,DutschV17} propose approximate solutions. 
An LSH for the discrete Fréchet distance is described in~\cite{driemel_locality-sensitive_2017}.
A follow-up paper~\cite{emiris2018} provides better theoretical approximation bounds using a slightly different approach, but their results do not apply to the setting that we focus on in this paper.
Sketches for the Hausdorff and discrete Fréchet distances are proposed in \cite{Astefanoaei18}, which gives an LSH scheme with similar properties of~\cite{driemel_locality-sensitive_2017}.

{\bf Verifying the Fréchet distance.} 
In order to improve the precision of the proposed LSH scheme, we suggest to filter the query results by verifying the distances for selected curves. However, verifying the distance is a non-trivial and expensive operation. It is known that the (discrete or continuous) Fr\'echet distance  between two fixed curves cannot be decided in strictly subquadratic time in the number of vertices of the curves, unless the Strong Exponential Time Hypothesis is false~\cite{Bringmann14}. 
The fastest algorithms for computing the continuous and discrete Fréchet distance are described in~\cite{buchin2014four} and~\cite{avraham2014}. Both algorithms take roughly quadratic time. 
However, \cite{DriemelHW12} shows that one can approximate the distance in near-linear time under certain realistic assumptions on the shape of the input curves. We use this algorithm to filter the query results, in order to improve the precision of our method.

\section{Preliminaries}\label{sec:prelim}

{\bf Continuous and discrete Fréchet distances}
A \emph{time series} (or \emph{trajectory}) is a series $(p_1,t_1),\ldots, (p_m,t_m)$ of measurements $p_i\in \Re^d$ of a signal taken at times $t_i$, where $0=t_1<t_2<\ldots <t_m=1$ and $m$ is finite. 
A time series denotes a \emph{polygonal curve} $p$ of length $m$  and defined by the sequence of \emph{vertices} $p_1,\ldots, p_m$.
A polygonal curve $p$ may be viewed as a continuous function $p: [0,n] \rightarrow \Re^d$ by linearly
interpolating $p_1,\dots,p_m$ in order of $t_i$, $i=1,\ldots, m$.
Each segment between $p_i$ and $p_{i+1}$ is called \emph{edge} $\overline{p_i p_{i+1}}=\{xp_i+(1-x)p_{i+1}|x\in [0,1]\}$.  
We let $|p|$ denote the \emph{length} of curve $p$, that is the number of vertices in $p$. 
The space of all polygonal curves in $\Re^d$ is denoted by $\Curves{d}$. 
As all our curves are polygonal, we omit the term ‘‘polygonal'' for the sake of simplicity.

For two vertices in $p,q\in \Re^d$, we let $d_E(p,q) = \|p-q\|_2$ denote their Euclidean distance.
Let $\Phi_n$ be the set of all continuous and non-decreasing functions $\phi$ from $[0,1]$ into $[1,n]$.  
The \emph{continuous Fréchet distance} of two curves $p$ and $q$, denoted by $d_F(p,q)$, is defined as 
\begin{align}
d_F(p,q) = \inf_{\substack{\phi_1\in \Phi_{|p|}\\ \phi_2\in \Phi_{|q|} }} \max_{t\in [0,1]} \left\|p_{\phi_1(t)} - q_{\phi_2(t)}\right\|_2.
\end{align}
Each pair    $(\phi_1,\phi_2)\in \Phi_{|p|} \times \Phi_{|q|}$ is called \emph{continuous traversal}, and it can been seen as a schedule for
simultaneously traversing the two curves,  starting on the first vertices of both curves at time~0 and ending on the last vertices at time~1.

The problem of verifying that the Fr\'echet distance between two curves is less than or equal to a threshold $r$ is usually done with the so-called \emph{free space diagram} \cite{AltGodau95}, which has quadratic cost in the worst case. 
However, it was shown in~\cite{DriemelHW12} that if the algorithm operates on simplified copies of the curves, then the complexity reduces to near-linear under certain assumptions on the shape of the curves. 
The simplification introduces an approximation error to the verification algorithm, but as shown in~\cite{DriemelHW12}, the error can be bounded if the simplification parameters are wisely chosen. 
By exploiting the bounded error, it is possible to use the simplification for  confirming or denying that two curves have distance at most $r$.

{\bf Range search and LSH.}
Given a set $S\subseteq \X$ of $n$ points in a domain $\X$, a distance function 
$d:\X \times \X \rightarrow [0,+\infty)$, and a radius $r>0$, 
the \emph{$r$-range search} (also known as range reporting) problem requires to construct a data structure that, for any given query point $q\in \X$, returns all points $p\in S$ such that $d(q,p)\leq r$.
We say that a point $p$ is a \emph{$r$-near} or \emph{$r$-far} point of $q$ if $
d(p,q)\leq r$  or $d(p,q)>r$, respectively;
if $r$ is clear in the context, we will just say that $p$ is a near or far point of $q$.

\emph{Locality Sensitive Hashing} (LSH) \cite{IndykM98} is a common tool for  $r$-range search in high dimensions.
For a given radius $r>0$ and approximation factor $c>1$, an LSH is an hash scheme $\mathcal H$ where for a random selected map $h\in \mathcal H$ and two points $x$ and $y$, we have that $\Pr_{h\in \mathcal H}[h(x) = h(y)] \geq p_1$ if $d(x, y) \leq  r$, and $\Pr_h[h(x) = h(y)] \leq p_2$ if $d(x, y) > c\cdot r$. 
Probabilities $p_1$ and $p_2$ depend on the LSH scheme and the quality of an LSH scheme is given by $\rho = \rho(H)  = \frac{\log 1/p_1}{\log 1/p_2}$ (values of $\rho$ closer to 0 are better).
Concatenation is a  technique for building an LSH scheme with a small collision probability $p_2$ of far points:
by concatenating $k\geq 1$ hash functions randomly and uniformly selected from $\mathcal H$, we get an LSH scheme with collision probability $p_1^k$ for near points and $p_2^k$ for far points.

The standard data structure based on LSH for solving the  $r$-range search problem is the following~\cite{IndykM98}. Assume that, after concatenation, we have $p_2\leq 1/n$.
Let $\ell_1,\ldots, \ell_L$ be $L$ functions randomly and uniformly chosen from $\mathcal H$.
The data structure consists of $L$ hash tables $H_1,\ldots H_L$: each hash table $H_i$ stores the input set $S$, partitioned by the hash function $\ell_i$.
For each query $q$, we compute the set $S_q =\cup_{i=1}^L H_i(\ell_i(q))$, where $H_i(\ell_i(q))$ denotes the set of points in $S$ colliding with $q$ under the hash function $\ell_i$.
Then, we scan $S_q$ and  remove all points with distance larger than $r$ from $q$; the remaining points are returned as $r$-near points of $q$.
If $L=\BT{p_1^{-1}}=\BT{n^\rho}$, then the above data structure returns in expectation a constant fraction of all near points of $q$.

\section{\fresh algorithm}\label{sec:algo}
We let $S$ denote our input set with $n$ curves of maximum length $m$, and let $q$ be a query curve.
For each query $q$, \fresh returns a set $O_q$ of pairs $(t,s_t)$ where $t\in S$ is a curve and $0 \leq s_t\leq 1$ is its \emph{score}.
Each score $s_t$ denotes the likelihood of $t$ to be close to the query $q$: a large value of $s_t$ implies a high probability that  $t$ is a $r$-near curve of $q$; further, if two curves $t$ and $t'$ have scores $s_t\leq s_{t'}$, then it is more likely that $t'$ is closer to $q$ than $t$. 
Curves with scores equal to $0$ are not reported since they are considered far from $q$. 

The above approach can generate both false negatives and false positives.
As we will later see, false negatives (i.e., near curves that are not reported in $O_q$) can be reduced by increasing the number of  LSH functions (i.e., the parameter $L$) used in the score computations.
On the other hand, false positives (i.e., far curves that are reported in $O_q$) can be reduced by verifying the  distance from $q$ of a subset of curves in $O_q$ with small scores.
Verifying that two curves have continuous Fréchet distance at most $r$ is however an expensive operation, we thus propose a heuristic based on a cascade of curve simplifications that efficiently rules out or confirms the distance between the curves. 

The section is organized as follows: Section~\ref{sec:scores} explains how scores are  computed; 
Section~\ref{sec:filtering} describes how to reduce false positives; 
Section~\ref{sec:frechetalgo}  shows how to verify if two curves have continuous Fréchet distance at most $r$.

\subsection{Score computations with LSH}\label{sec:scores}
At a high level, the score $s_p$ of a curve $p \in S$ with query $q$  is given by the normalized number of collisions with $q$ under $L\geq 1$  hash functions from the
 LSH scheme  $\mathcal G_\delta^k$ described below, where  $\delta$ and $k$ are suitable parameters. 

{\bf LSH scheme $\mathcal G_\delta^k$.}  
Our starting point is the LSH scheme $\mathcal{\hat G}_\delta$ in~\cite{driemel_locality-sensitive_2017}, which maps each curve into a smaller curve with vertices from a random shifted grid
\[
G_{\delta,t} = \brc{ (x_1,\dots,x_d) \in \Re^d ~|~ \forall~ i\in [d]  ~\exists~ j \in \Na:~  x_i = j \cdot \delta +t}
\]
where $\delta>0$ is the side of the grid and $t=(t_1,\ldots t_d)$ is a random variable uniformly distributed in $[0,\delta)^d$.
For a curve $p$ with vertices $p_1,\dots,p_{m}$, the function $g_{\delta,t}(p)$ returns the curve obtained by: 1) replacing each vertex $p_i$ with its closest grid vertex  in $G_{\delta,t}$; 2) removing consecutive duplicates in the new curve.
The LSH family $\mathcal{\hat G}_\delta$ is defined as $\mathcal{\hat G}_\delta=\{g_{\delta,t}, \forall t\in [0,\delta)^d\}$.
We also define  $\mathcal{\hat G}^k_\delta$ as the LSH family obtained by concatenating $k\geq 1$ copies of hash functions uniformly and independently selected in $\mathcal{\hat G}_\delta$.
We  have that $\Pr_{g^k \in \mathcal {\hat G}^k_{\delta}}[g^k(q)=g^k(p)]= \Pr_{g\in \mathcal{\hat G}_{\delta}}[g(q)=g(p)]^k$: the lower collision probability of far curves allows to decrease false positives.

\fresh requires the computation of a large number  of hash values in $\mathcal{\hat G}^{k}_\delta$: indeed, $k \cdot L \cdot n$ hash values are computed at construction time and  $k\cdot L$ hash values for each query.
We speed up the hash computation  with the tensoring approach. Tensoring was initially proposed in~\cite{andoni_efficient_2006} and then further studied in~\cite{christiani_fast_2017}; to the best of our knowledge, it has only been used in practice in~\cite{SundaramTSMIMD13}.
The tensoring approach generates $L$ hash functions building on two collections of  $\sqrt{L}$ hash functions, reducing the actual number of hash computations  by a   $\sqrt{L}$ factor.
Specifically, let $\Lambda_1=\{g_1,\ldots, g_{L'}\}$ and $\Lambda_2=\{g'_1,\ldots, g'_{L'}\}$ be two groups of $L'=\sqrt{L}$ random hash functions from $\mathcal{\hat G}_{\delta}^{k/2}$.
Then, it is possible to construct $L'\cdot L'=L$ LSH hash functions from $G_\delta^k$ by concatenating the pair $(g_i,g'_j)$ for all possible values of $i$ and $j$ in $\{1,\ldots L'\}$.
This technique reduces the number of hash value computations for the initial data structure construction from $k \cdot L \cdot n$ to $k \cdot \sqrt{L}\cdot n$, and for the query procedure from $k\cdot  L$ to $k \cdot \sqrt{L}$.

Finally, as storing and searching signatures is quite inefficient,  we map all signatures on integers with  the multiply-shift hashing scheme $\mathcal H$ in~\cite{Dietzfelbinger97}.
We denote with $\mathcal G_{\delta}^{k}$ the LSH hash family obtained by first using the tensoring approach to construct (a subset of) $\mathcal{\hat G}_{\delta}^{k}$, and then by applying the multiply-shift hashing $\mathcal H$ on the signature.
We observe that the signature of a curve does not need to be generated and stored: while we scan a curve $p$ to compute its signature, the hash value $h(g(p))$ is built on the fly.

{\bf Data structure.} The data structure of \fresh for efficiently computing the scores leverages on the traditional approach for solving range search  with LSH.
$L\geq 1$ hash functions $g_1,\ldots, g_L$ are randomly chosen from the above LSH family $\mathcal G_\delta^k$, for suitable values of $\delta$ and $k$; then for each $g_i$, a hash table $H_i$ is created for storing the $n$ input curves partitioned by  $g_i$.
For each query $q$, we compute the multiset $T_q = \cup_{i=1}^L H_i(g_i(q))$, where $H_i(g_i(q))$ denotes the set of curves colliding with $q$ under $g_i$. 
If $t\in T_q$ and its multiplicity in $T_q$ is $\hat s_t$, then its score $s_t$ is $\hat s_t/L$.
Note that the hash tables do not need to store the complete curves but just their identifiers: thus, the space required by the data structure is $I+\BT{Ln}$ memory words, where $I$ is the number of words to store $S$.

\subsection{Filtering false positives}\label{sec:filtering}
All curves with non-zero score are not too far from the query: indeed, if the hash function uses a grid of side length $\delta$, then all colliding curves have maximum distance $\delta$.
However, as in general $\delta > r$ (in our experiments $\delta = 4dr$, where $d$ is the point dimension), we may report some  curves with distance in $(r,\delta]$.
To improve the precision, a simple approach is to set a threshold $\Delta$ and verify all curves with scores less than $\Delta$.
However, the limitations of this approach are: 1) it is not clear how to select the best $\Delta$ as it might be query dependent; 2) $\Delta$ does not directly allow to trade precision and running time.
The approach used in \fresh is to verify a fraction $\tau$, with $0\leq \tau \leq 1$, of the curves in $O_q$ with smaller scores. 
The parameter $\tau$ can  be used for trading  performance (with $\tau=0$ no curve in $O_q$ is verified) with precision (with $\tau=1$, all curves in $O_q$ are verified which implies a 100\% precision).

\subsection{Verifying the Fréchet distance}\label{sec:frechetalgo}
Verifying that two curves $p$ and $q$ are within Fréchet distance $r$ is an expensive operation~\cite{Bringmann14}:
to speed up this operation, we introduce the procedure \textsc{Verify} for checking if two curves $p$ and $q$ have continuous Fréchet distance less than or equal to $r$.
\textsc{Verify}  consists of two procedures, named  \textsc{VerifySimpl} and \textsc{VerifyHeur}, that exploit strategies from~\cite{DriemelHW12,BaldusB2017,BuchinDDM17}: each procedure can successfully show that $d_F(p,q) \le r$ or $d_F(p,q) > r$, or it can fail and do not provide an answer.
Procedure \textsc{VerifyHeur} exploits the heuristics \emph{Equal-time alignment}~\cite{BuchinDDM17}, \emph{Greedy algorithm}~\cite{BaldusB2017} and \emph{Negative filter}~\cite{BaldusB2017}, and it stops as soon as one of them succeeds. 
On the other hand, procedure \textsc{VerifySimpl} is a  decision procedure based on the concept of simplification in \cite{DriemelHW12}: $p$ and $q$ are mapped on suitable smaller trajectories $p'$ and $q'$ through a transformation based on a parameter $\epsilon\geq 0$ ($\epsilon=0$ gives the original curves).
Evaluating distance predicates on $p'$ and $q'$ allows to answer distance predicates on $p$ and $q$, by suitable setting the parameter $\epsilon$.

Procedure \textsc{VerifyHeur} is the application of the following heuristics, stopping as soon as one of them succeeds.
\begin{itemize}
\item \emph{Equal-time alignment}~\cite{BuchinDDM17}. 
  This heuristic performs a traversal of the two curves moving at the same speed on both, providing an upper bound to the Fréchet distance.
  If we define $\Phi_{x}(t) = t x$, this heuristic verifies
  \[
    \max_{t \in [0, 1]} || p_{\Phi_{|p|}(t)} - q_{\Phi_{|q|}(t)} ||_2\le r
  \]
  which can be done in linear time.
\item \emph{Greedy algorithm}~\cite{BaldusB2017}.
It provides an upper bound on the continuous Fréchet distance by finding an alignment with a greedy approach. 
We construct the following traversal of $p$ and $q$: 
1) $p_1$ and $q_1$ are matched; 
2) after matching vertices $p_i$ and $q_j$, we match $p_{i'}$ and $q_{j'}$, for $(i',j')\in\{(i+ 1, j),(i, j + 1),(i+ 1, j + 1)\}$ minimizing $\|p'_{i'}-q_{j'}\|_2$.
We ignore from these three options the ones that would make $i > |p|$ or $j > |q|$.
If during the whole traversal we stayed at distance $\le r$, we can conclude that $p$ and $q$ are $r$-near.
\item \emph{Negative filter}~\cite{BaldusB2017}. 
  This heuristic seeks to prove that, for some vertices of $p$, there are no vertices of $q$ within distance $r$ they can be aligned to,
  providing a certificate that the two curves are at distance greater than $r$.
  For each vertex $p_j$ of $p$, we define $q^\leftarrow_{p_j}$ as the first vertex of $q$ that can be aligned with $p_j$.
  For this to be possible, such a vertex needs to be within distance $r$ from $p_j$, and needs to appear on $q$ after vertex $q^\leftarrow_{p_{j-1}}$, because of the definition of Fréchet distance.
  Since the first vertex of $p$ has to be aligned with the first vertex of $q$, we have that $q^\leftarrow_{p_1}=q_1$.
  Then, for $j \in [2, |p|]$ the heuristic proceeds in trying to define $q^\leftarrow_{p_j}$. If for some $j$ this is not possible, then $p$ and $q$ are farther than $r$.
  This heuristic is not symmetric, therefore we can apply it two times swapping arguments.
\item \emph{Full verify}. 
  If all of the above heuristics fail to verify the distance, we apply the exact algorithm in~\cite{AltGodau95} based on free space diagram.
\end{itemize}

To further speedup the verification of a pair of curves $p$ and $q$, we also adopt the decision procedure proposed in~\cite[Lemma 3.6]{DriemelHW12}, which we deem here \textsc{VerifySimpl}.
This scheme is based on the concept of $\mu$-simplification (also presented in~\cite{DriemelHW12}), constructed as follows for a curve $p$ and $\mu > 0$.
First mark $p_1$ and set it as the current vertex.
Then, scan the curve from the current vertex until we reach the first $p_j$ such that $||p_j - p_1||_2 > \mu$: we mark $p_j$ and set it as the current vertex.
The procedure is repeated until we reach the last vertex, which is marked as well.
The marked vertices make up the simplified curve, which is denoted with $\simpl{p, \mu}$ and is computed in linear time.
The decision scheme builds simplifications of $p$ and $q$, controlled by a parameter $\epsilon > 0$.
Let $r'=r/(1+\epsilon/3)$.
Define $\mu^-=r\epsilon/28$ and $\mu^+=r\epsilon/(28 \cdot(1+\epsilon/3))$, and let 
\[
  r^- = r\cdot\left(1 + \epsilon/14\right)
  \quad
  \mbox{and}
  \quad
  r^+ = r\cdot\left(\frac{3(1+\epsilon/14)}{3 + \epsilon}\right)
\]
note that $r^+ < r^-$.
First, we verify with \textsc{VerifyHeur} if 
\[d_F(\simpl{p, \mu^-}, \simpl{q, \mu^-}) > r^-\] 
If this is the case, the procedure reports that $d_F(p,q)> r$.
Otherwise, we further verify with \textsc{VerifyHeur} if 
\[d_F(\simpl{p, \mu^+}, \simpl{q, \mu^+}) \le r^+\]
If the answer is affirmative, we report $d_F(p,q) \le r$.
It may be that neither of the two checks gives a positive answer.
In this case, the procedure reports that it cannot give an answer.

Procedure \textsc{Verify} is then the following:
\begin{enumerate}
\item In the first stage, we only consider the first ($p_1$ and $q_1$) and last vertices ($p_{|p|}$ and $q_{|q|}$) of $p$ and $q$.
If $||p_1-q_1||_2>r$ or $||p_{|P|}- q_{|Q|}||_2>r$, then the two curves cannot be $r$-near by the definition of continuous traversal.
We call this heuristic \textsc{Endpoints}.

\item In the second stage, we look at the bounding boxes of the two curves. If the $\ell_1$ distance of corresponding corners of the bounding boxes is larger than $r$, then the two curves cannot be $r$-near~\cite{DutschV17}. We call this heuristic \textsc{BoundingBox}.

\item In the third stage, we use \textsc{VerifySimpl} with decreasing values of $\epsilon$ (which will be fixed in the experimental analysis), corresponding to simplifications becoming less aggressive.
For a given $\epsilon$, if \textsc{VerifySimpl} can give an answer, then we return it, otherwise we move to the next $\epsilon$.

\item The fourth stage runs if none of the calls to \textsc{VerifySimpl} could return an answer: in this case we return the result of the invocation of \textsc{VerifyHeur} on the original curve.
\end{enumerate}

\section{Experimental evaluation}\label{sec:exp}
In this section, we present our experimental evaluation of \fresh. 
Section \ref{sec:setup} describes the  setup of our experiments, including the benchmarks and the exact baseline algorithm used as reference.
Section \ref{sec:lshclassifier} analyzes the performance and quality of the LSH scheme in \fresh, without the partial verification to reduce false positives: in particular, we investigate how the number of LSH repetitions ($L$) and of LSH concatenations ($k$) affect performance and quality  (recall/precision).
Section \ref{sec:exp:partial-verification} examines how the partial verification affects the performance and precision under different values of the fraction $\tau$ of verified candidate curves, and it analyses the effectiveness of the various heuristics used in \fresh  to prune false positives.

\subsection{Experimental setup}\label{sec:setup}
\paragraph{\bf Hardware} 
We implement our algorithm in C++ with OpenMP, using the \texttt{gcc} compiler
version 4.9.2. We run the experiments on a Debian GNU/Linux machine (kernel
version 3.16.0) equipped with 24GB of RAM, and an Intel I7 Nehalem processor
(clock frequency 3.07GHz). 

\paragraph{\bf Datasets}

As benchmarks we use datasets from the UCR collection~\cite{UCRArchive}, which is comprised of 85 datasets of trajectories in one dimension.
For brevity, we report on the 7 largest datasets of this collection.
We also include in our benchmark a dataset of road trips in San Francisco that was used in the SIGSPATIAL 2017 challenge~\cite{werner2018acm}, along with the TDrive dataset~\cite{t-drive-dataset}.
Both are datasets of trajectories in 2 dimensions.

\begin{figure*}[t]
  \includegraphics[width=\textwidth]{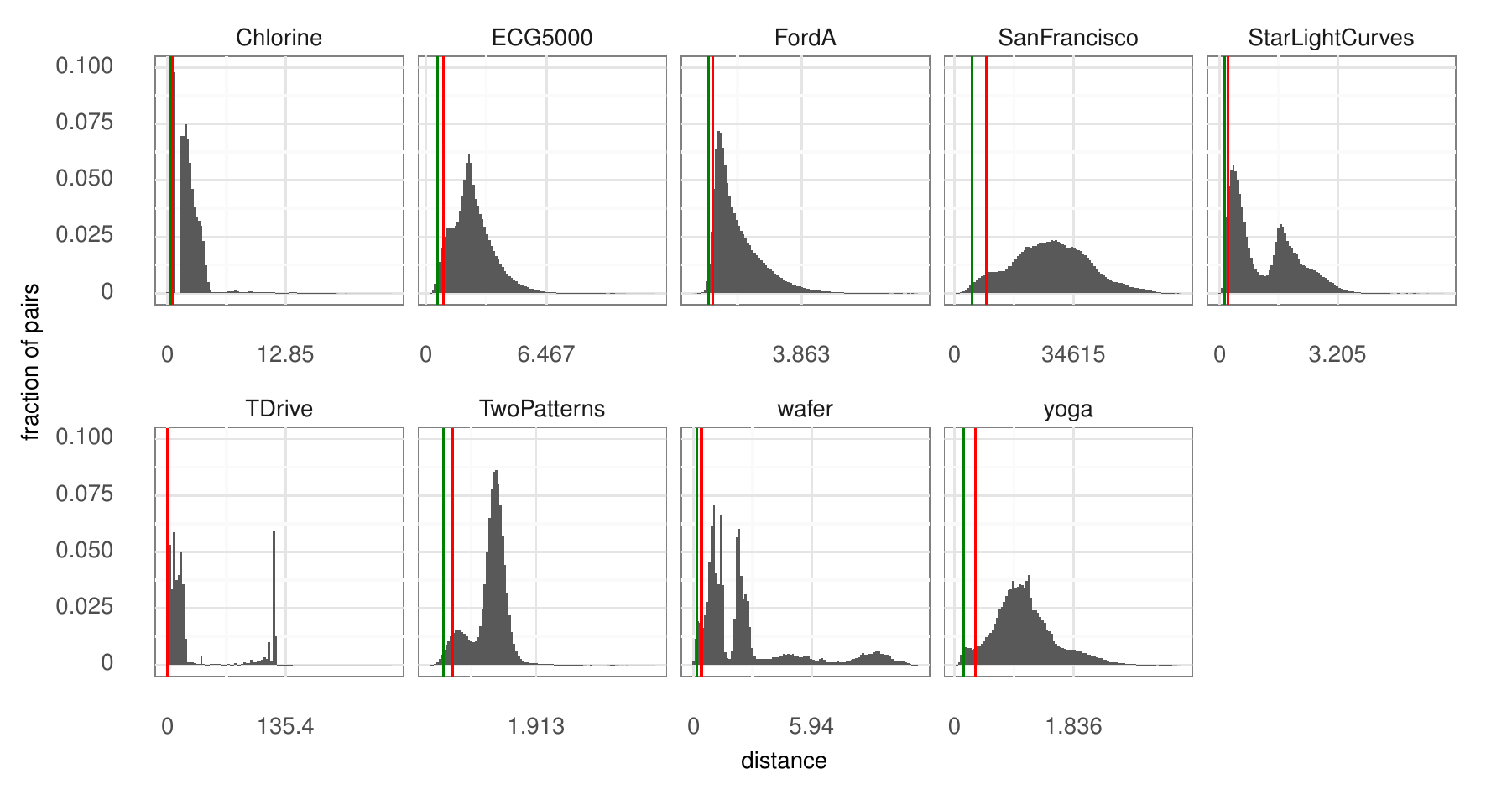}
  \caption{Distribution of pairwise distances for all the datasets considered.
    The green line highlights the first percentile, the red one highlights the fifth percentile.
    \label{fig:distance-distribution}}
\end{figure*}

For each dataset, we perform a self-similarity join using a set of fixed Fréchet distance thresholds, by solving the $r$-range search problem for each curve of the dataset.
The thresholds are set to the first and fifth percentiles of the pairwise distances for any given dataset, so that the output size is 1\% and 5\% of the number of possible pairs, respectively.
Given the large number of possible pairs, these percentiles are computed on the pairwise distances of a sample of 1000 points of each dataset.
Figure~\ref{fig:distance-distribution} gives the distribution of pairwise distances in the datasets we are considering.
Each result is the average over at least 5 runs.

\begin{table}
  \newcommand{\dutschmarker}{~$\star$}
  \newcommand{\baldusmarker}{~$\ddagger$}
    \begin{center}
      \small
      \begin{tabular}{llr}
      \toprule
      dataset & range & best time \\
      \midrule
      Chlorine & 0.34 (first) &     24\dutschmarker{} \\
           & 0.52 (fifth) &     91\dutschmarker{} \\
      ECG5000 & 0.62 (first) &     29\dutschmarker{} \\
           & 0.92 (fifth) &    102\dutschmarker{} \\
      FordA & 1.07 (first) &    299\dutschmarker{} \\
           & 1.20 (fifth) &   1190\dutschmarker{} \\
      yoga & 0.14 (first) &     23\dutschmarker{} \\
           & 0.33 (fifth) &     87\dutschmarker{} \\
      SanFrancisco & 5213.21 (first) &    413\dutschmarker{} \\
           & 9205.43 (fifth) &    417\baldusmarker{} \\
      StarLightCurves & 0.13 (first) &    548\dutschmarker{} \\
           & 0.21 (fifth) &   2949\dutschmarker{} \\
      TDrive & 0.17 (first) &   3913\dutschmarker{} \\
           & 0.23 (fifth) &  20372\dutschmarker{} \\
      TwoPatterns & 0.56 (first) &     76\dutschmarker{} \\
           & 0.68 (fifth) &    121\dutschmarker{} \\
      wafer & 0.14 (first) &     70\dutschmarker{} \\
           & 0.39 (fifth) &    134\dutschmarker{} \\
      \bottomrule
      \end{tabular}
    \end{center}
    \captionof{table}{Baseline times (in seconds) for the two different radii, which are defined, respectively, as the first and fifth percentile of all pairwise distances.
    Results marked with \baldusmarker{} were obtained using the code by Baldus et al.~\cite{BaldusB2017},
    the ones marked with \dutschmarker{} were obtained using the code by Dutsch et al.~\cite{DutschV17}.
  \label{tab:baseline}}
\end{table}

\paragraph{\bf Baseline}

To establish a baseline, we ran the code provided by the three winners of the SIGSPATIAL 2017 challenge~\cite{BaldusB2017,BuchinDDM17,DutschV17}, compiled with all optimizations enabled and ran with
the default parameters.
Table~\ref{tab:baseline} reports these results.

\begin{figure*}[t]
    \includegraphics[width=\textwidth]{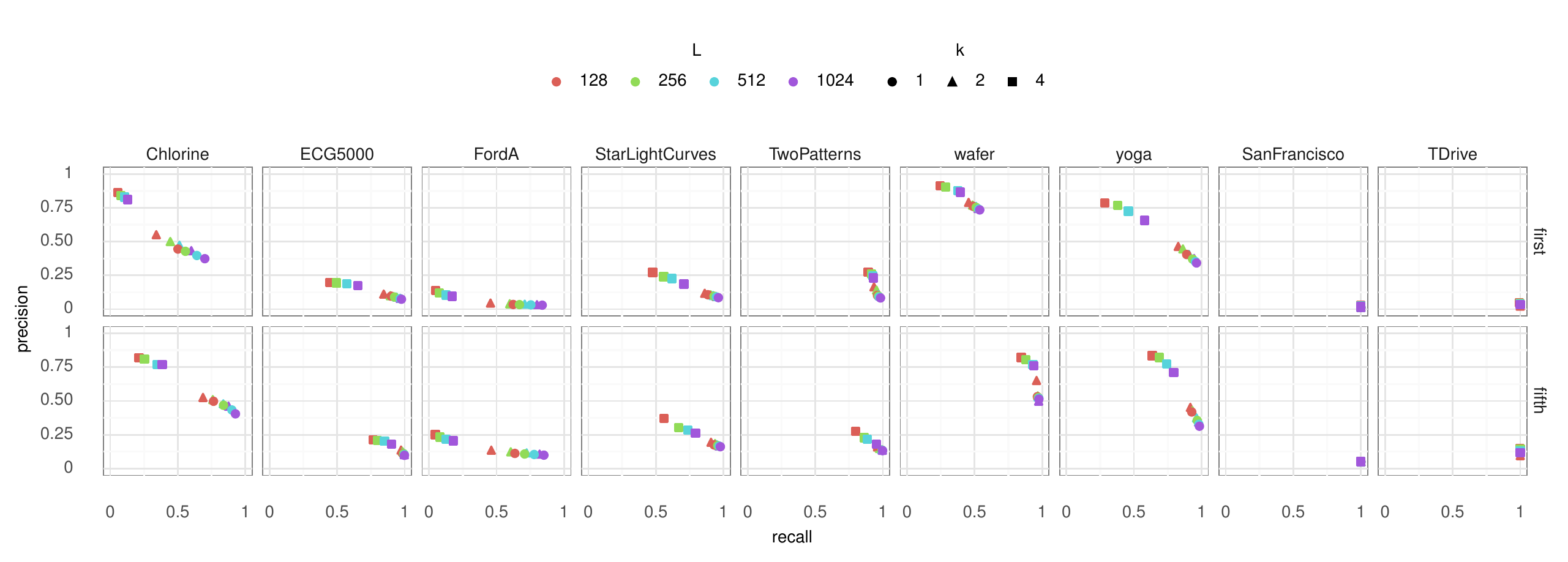}
    \caption{Performance in terms of precision and recall of \fresh  on all the datasets considered. The color of a point  denotes the number of repetitions $L$, while the shape of a point represents the number of concatenations $k$.\label{fig:precision-recall}}
\end{figure*}

\subsection{Evaluating the LSH scheme}
\label{sec:lshclassifier}
\begin{figure*}[t]
  \begin{minipage}{.73\textwidth}
    \centering
    \includegraphics[width=\columnwidth]{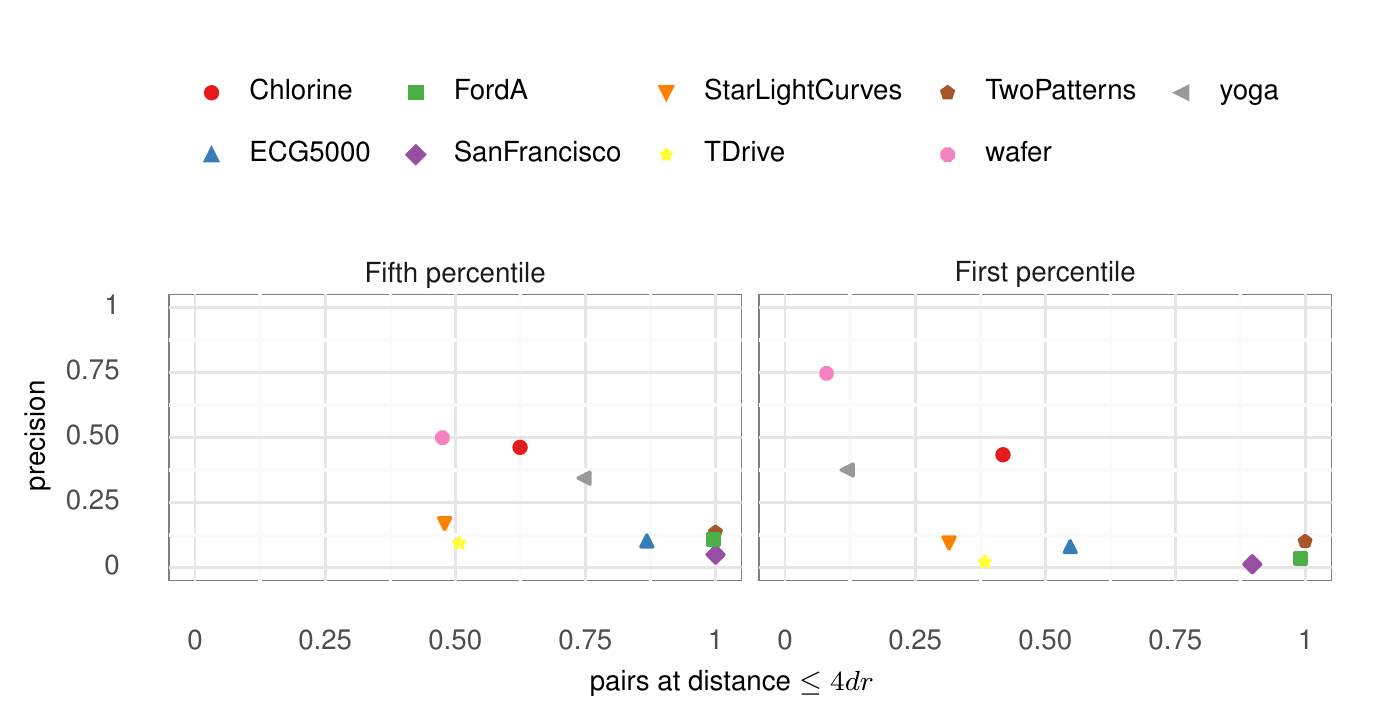}
  \end{minipage}
  \begin{minipage}{.25\textwidth}
    \caption{Fraction of pairs below $4dr$ versus precision, for $k=2$, and $L=1024$.
      On dataset where such a fraction is high, the precision of the LSH scheme tends to be low.
    \label{fig:mass}}
  \end{minipage}
\end{figure*}

\begin{figure*}[t]
  \includegraphics[width=\textwidth]{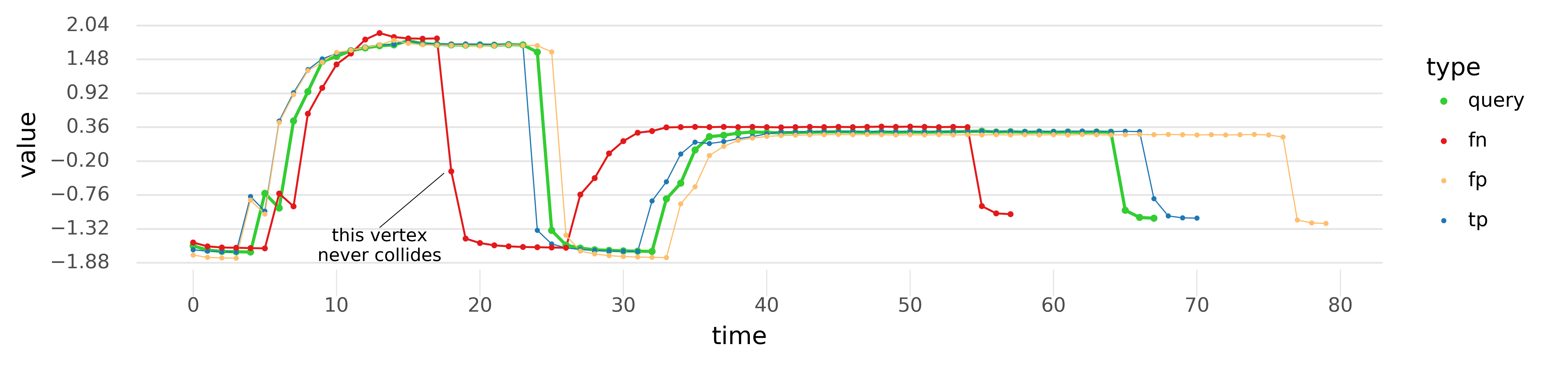}
  \caption{Curve 0 of the wafer dataset as a query (green) for $r=0.14$, $k=2$ and $L=1024$, in the context of relevant curves with respect to the LSH scheme: false positives (orange), false negatives (red) and true positives (blue).
  The spacing of the grid along the value axis is equals to $4r$, which is the size of the grid used by the LSH scheme for building signatures.\label{fig:wafer}}
\end{figure*}

We analyze how the LSH scheme affects the performance and quality of \fresh without the partial verification. 
In other words, each pair colliding in at least one of the $L$ repetitions (i.e., with a non-zero score) is reported as a positive match, without further verification.
We test this setup using hash values obtained as the concatenation of $k=1,2,4$ hash functions and with $L=128,256,512,1024$ repetitions, setting the grid size to $\delta=4dr$.
Figure~\ref{fig:precision-recall} reports, for each dataset and combination of parameters, the performance in the precision-recall space.
The recall is the fraction of true positives reported by the algorithm over all the positives in the ground truth, whereas the precision is the fraction of true positives over the predicted positives (i.e., the sum of true positives and false positives).
Both scores range from 0 to 1, with 1 being the best, hence in the plots of Figure~\ref{fig:precision-recall} we have that the closer the top right corner, the better the performance.
Note that we use the precision instead of the false positive rate due to the large number of negatives in the ground truth, which makes very easy to attain a small false positive rate.

In general, we have that increasing the number of repetitions $L$ improves the recall, lowering the precision, as expected.
Symmetrically, increasing $k$ makes the LSH more selective, hence it increases the precision, at the expense of the recall.
Note that on some datasets our LSH technique is more effective than on others.
In general, using sufficiently many repetitions we can get good recall, while getting a good precision is harder, and may be very costly in terms of recall.
We will address this problem in the next subsection.

On the SanFrancisco and TDrive datasets we get perfect recall and low precision, almost irrespective of the configuration of parameters.
This is due to the distance distribution of these datasets: by setting the query range to the first and fifth percentiles of distances, the algorithm constructs grids with a resolution so large that almost all curves collide with the queries.

\begin{figure*}[t]
    \includegraphics[width=\textwidth]{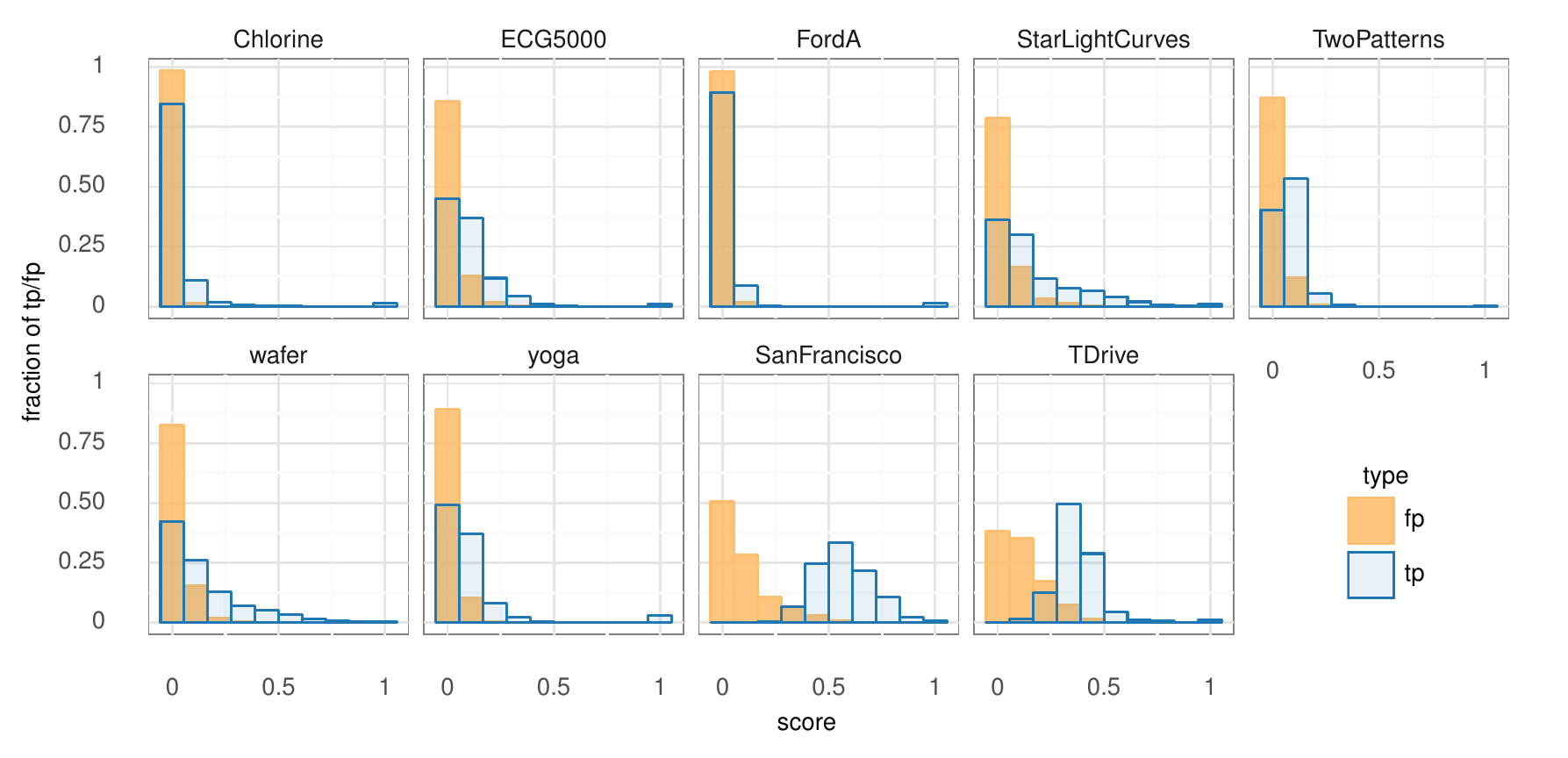}
    \caption{
      The distribution of scores assigned to colliding pairs for $k=2$ and $L=1024$, with query radius equals to the first percentile of distances, shows that the majority of false positive  pairs (fp, in orange)  have lower scores than the true positive colliding pairs (tp, in blue), with some overlapping of the two distributions.      
      The results for other configurations of parameters are similar.
      Note that in this plot each orange (resp. blue) bar is scaled with respect to the total number of false positives (resp. true positives) and not the total number of colliding pairs: this is to appreciate the overall distribution.
    \label{fig:score-distribution}}
\end{figure*}

Among the others, the wafer dataset deserves a particular attention.
For the query range equals to the first percentile of the pairwise distances, Figure~\ref{fig:precision-recall} shows that the recall is just slightly above 0.5 at best.
While a low precision can be fixed for all datasets, as we shall see in the next subsection, the recall on wafer seems resistant to increases of $L$.
To understand why this happens, we can look at the behavior of a single query, as reported in Figure~\ref{fig:wafer}.
Along with the one-dimensional query curve itself, we plot two curves that collide with the query under the LSH scheme, one false positive and one true positive, and a curve that did not collide but should have, i.e. a false negative.
In terms of recall, the false negatives are the relevant curves to look at: having zero false negatives implies a perfect recall.
Therefore, the poor performance on wafer is due to the fact that many curves are classified as being far from the query when they are actually close, which happens if the misclassified curve and the query do not collide in any of the $L$ repetitions.
Looking at Figure~\ref{fig:wafer} we can see why this happens.
The query (green curve), has a sudden jump downward around time 25, with no vertices in the segment connecting the extremes of the jump.
The false negative curve (in red) has a similar jump around time 18.
However, in this case, there is one vertex between the extremes of the jump.
Under the LSH scheme described in Section~\ref{sec:algo}, two curves collide (and hence have a non-zero score) only if they have the same signature, which is computed by snapping vertices to a randomly shifted grid of resolution $4dr$, i.e. $4r$ for one-dimensional dataset such as wafer.
The grid of Figure~\ref{fig:wafer} has a resolution $4r$ along the value axis.
It is clear that, no matter the random shift of the grid, the point of the red curve in the middle of the jump will never snap to the same grid line as any point of the green curve in the analogous jump, because no such point exists.

A simple solution to this problem is to add more vertices to the curves, by interpolation, in the jumps.
This preprocessing does not change the Fréchet distance between any two curves.

\subsection{Improving the precision by partial verification}
\label{sec:exp:partial-verification}

In this section we verify the trade-off between precision and running time proposed in Section~\ref{sec:filtering}.
From the previous experiments we selected a configuration of parameters striking a good balance of recall and precision on most datasets: $k=2$ and $L=1024$.
For $\tau\in\{0, 0.1, 0.2, 0.5, 1\}$ we run the algorithm evaluating the $\tau m$ pairs with lowest non-zero scores, where $m$ is the number of pairs with non-zero scores.
When $\tau=0$, the algorithm  runs in the same configuration used in the previous subsection, when $\tau=1$ the algorithm verifies all the colliding pairs.
We apply 3 simplifications in the verification pipeline, using $\epsilon=10,1,0.1$, from coarsest to finest.

First, we consider the distribution of scores before any verification happens, to assert that verifying the lowest-score pairs is actually sound (Figure~\ref{fig:score-distribution}).
We have that the false positive pairs (colored in orange)  have lower scores than the true positive colliding pairs (in blue), with some overlapping of the two distributions.
Therefore, verifying pairs starting from the low-score ones seems like a sensible choice, since we are likely to get rid of many false positives, which we expect to improve the recall.
Note that verifying some pairs does not remove true positives (neither it can introduce them), therefore the recall remains unchanged, irrespective to the fraction of pairs $\tau$ that we verify.

\begin{figure*}[t]
    \includegraphics[width=\textwidth]{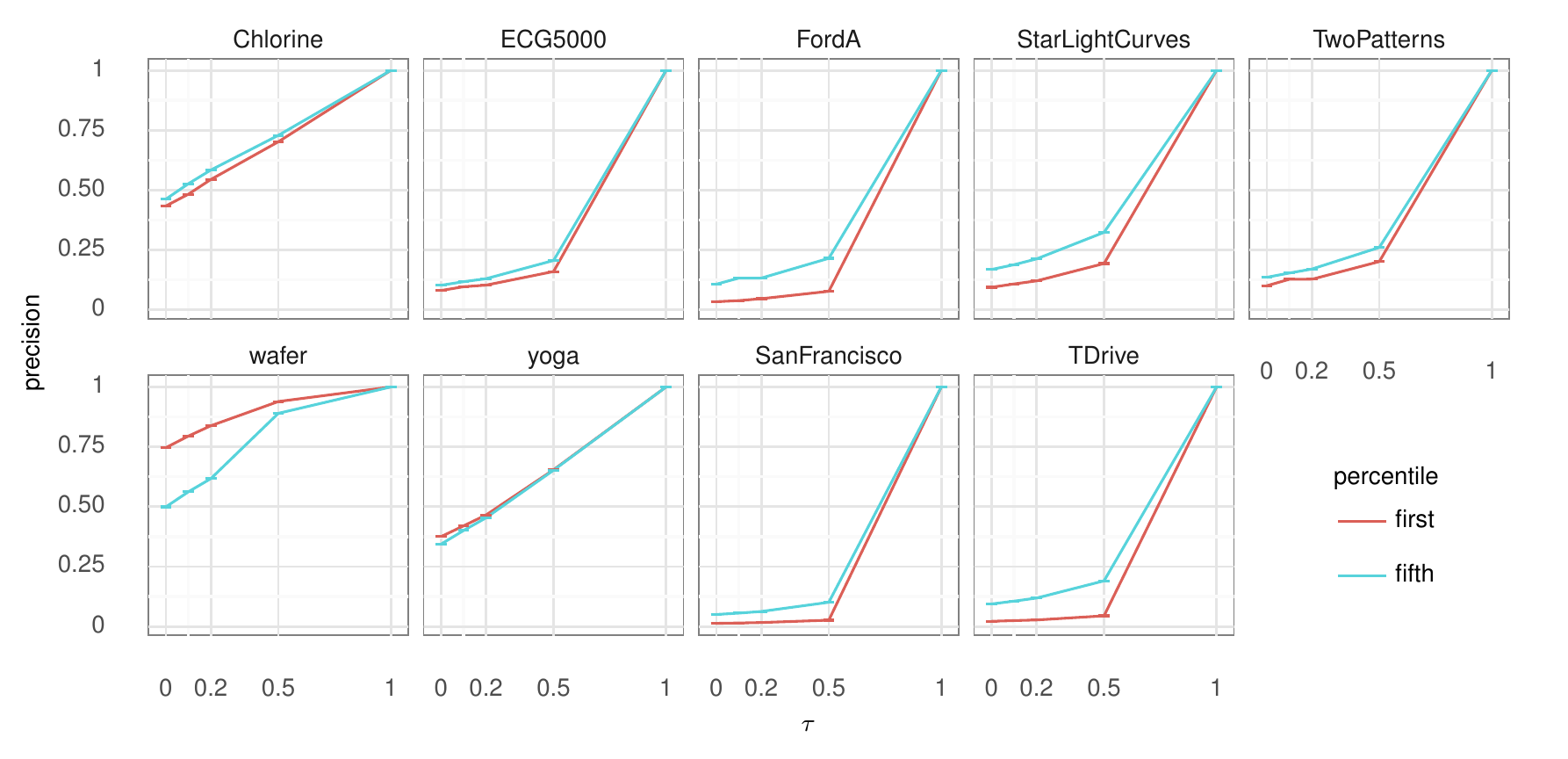}
    \includegraphics[width=\textwidth]{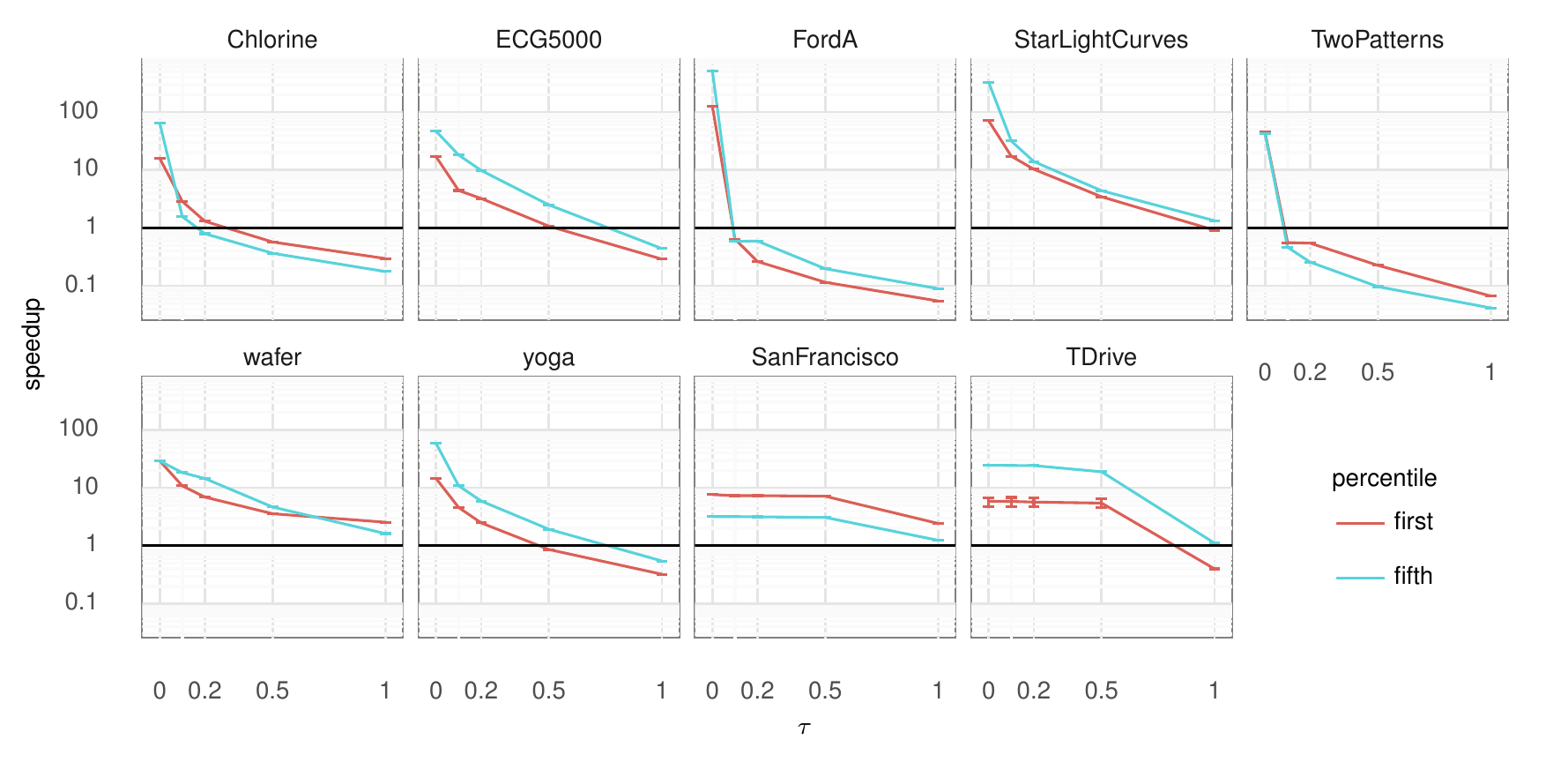}
  \caption{
    Precision and speedup per pair given for varying $\tau$, for $k=2$, $L=1024$.
    The black line on the speedup plots marks speedup 1, i.e. the performance of the best baseline algorithm.
    \label{fig:partial-evaluation}}
\end{figure*}

We now move to assess the influence of the fraction of verified pairs $\tau$ on the precision and the runtime performance (Figure~\ref{fig:partial-evaluation}).
For measuring the latter, we focus on the  \emph{speedup}, defined as the ratio between the time of the baseline and LSH based algorithm.
As we expect, increasing $\tau$ increases the precision, with perfect precision when $\tau=1$, when all the pairs are verified and the algorithm reports no false positives.
The speedup decreases with the increase of $\tau$: this is because we evaluate more and more pairs, which is a costly operation.
We observe that on two-dimensional trajectories, the speedup that can be obtained is larger than on one-dimensional datasets, even at higher precision values.

\begin{figure*}[t]
  \includegraphics[width=\textwidth]{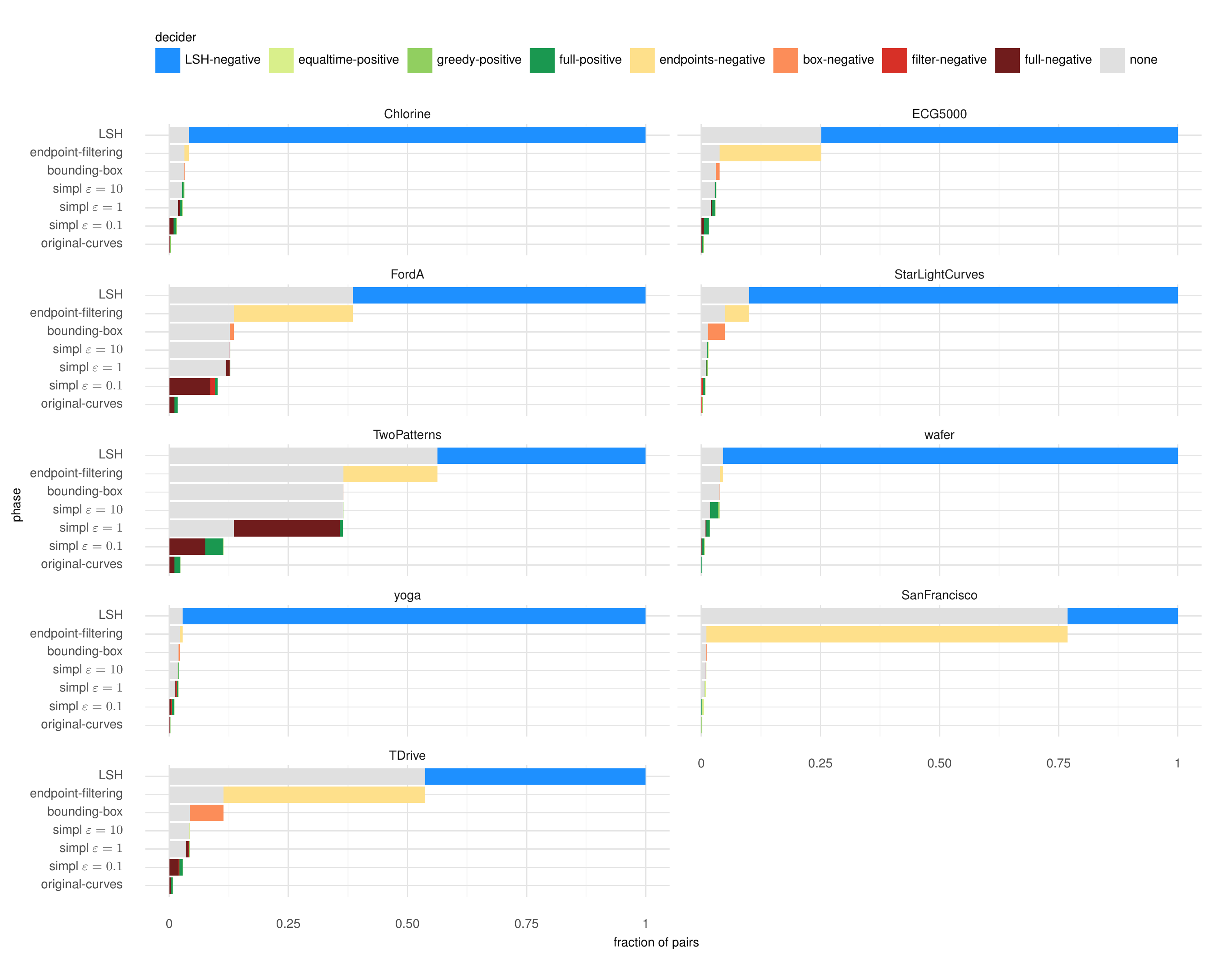}
  \caption{
    Breakdown of the effect of the various heuristics used to decide whether a pair is a positive match or not.
    The hue of the colors increases with the cost of the heuristic, so \emph{full-negative} is more expensive to compute than \emph{endpoints-negative}.
    \label{fig:breakdown}}
\end{figure*}
Finally, we analyze the contribution to the decision process of the LSH and the various heuristics employed (Figure~\ref{fig:breakdown}).
We concentrate on a single run, for each dataset, with $k=2$, $L=1024$ and the radius set to the first percentile of distances, evaluating all pairs with nonzero score.
The parts shaded in gray denote pairs for which the algorithm was not able to reach a decision and needed to move to the next stage.
Then, parts in shades of green (resp. red) denote pairs for which a positive
(resp. negative) decision was reached using one of the heuristics.
The pairs excluded by the LSH scheme are shaded in blue rather than red, to remark that even if they are rejected as negatives they may contain some false negatives: the larger the blue bar, the more effective the filtering power of the LSH scheme.
Some datasets are more amenable to be processed with the LSH strategy, and this is in line with the precision results reported in Figure~\ref{fig:precision-recall}.
Of the pairs surviving this first filtering, several can be discarded by looking at the endpoints, as shown by the \emph{endpoint-filtering} column in the plot.
The simplifications have varying degrees of effectiveness, depending on the dataset: on some datasets coarser simplifications are effective, whereas on some others we have to  use finer simplifications (i.e., with a smaller $\epsilon$).

\section{Conclusion}\label{concl}
As future work, it would be interesting to develop a general approach that merges the techniques in \fresh with the ones used in the exact solutions of the ACM SIGSPATIAL competition;
more generally, a challenge is understanding which input features make a solution more efficient than others.
The filtering approach used in \fresh can be enriched by using techniques for classifier assessment that consider the different costs that false positives and false negatives can have on the final application.
Finally, we observe that the LSH scheme for the discrete Fréchet distance in \cite{driemel_locality-sensitive_2017} also holds under the DTW distance: an interesting direction is to extend and analyze \fresh to report near curves under the DTW distance and other distance measures.

{\bf Acknowledgments.}
The authors would like to thank M. Aum\"{u}ller, K. Bringmann, F. D\"{u}tsch, R. Pagh  and J. Vahrenhold for useful comments, and the developers of the UCR collection.
This work has been partially supported by: ERC project “Scalable Similarity Search”, NWO Veni project 10019853, SID 2018 and 2017 projects of the University of Padova.

\bibliographystyle{plain}
\balance
\bibliography{biblio}

\end{document}